\documentclass[aps,prb,reprint,superscriptaddress]{revtex4-1}
\usepackage{graphicx}
\usepackage{epstopdf}
\usepackage{longtable}
\begin{document}

\title{Descriptors for thermal expansion in solids}
\author{Joseph T. Schick}
\email[]{joseph.schick@villanova.edu}
\affiliation{Department of Physics, Villanova University, Villanova, PA 19085-1478, USA}

\author{Abhijith M. Gopakumar}
\affiliation{Department of Chemistry, University of Pennsylvania, Philadelphia, PA 19104-6323, USA}

\author{Andrew M. Rappe}
\affiliation{Department of Chemistry, University of Pennsylvania, Philadelphia, PA 19104-6323, USA}

\begin{abstract}
Thermal expansion in materials can be accurately modeled with careful 
anharmonic phonon calculations within density functional theory.  
However, because of interest
in controlling thermal expansion and the time consumed evaluating thermal
expansion properties of candidate materials, either theoretically or 
experimentally, an approach to rapidly identifying materials with desirable
thermal expansion properties would be of great utility.
When the ionic bonding is important in
a material, we show that the fraction of crystal volume occupied by ions, 
(based upon ionic radii), the mean bond coordination, and the deviation of 
bond coordination are descriptors that  
correlate with the room-temperature coefficient of 
thermal expansion for these 
materials found in widely accessible databases.
Correlation is greatly improved by combining these descriptors
in a multi-dimensional fit. 
This fit reinforces the physical interpretation that 
open space combined with
low mean coordination and a variety of local bond coordinations
leads to materials with lower coefficients of thermal 
expansion, materials with single-valued local coordination and
less open space have the highest coefficients of thermal expansion. 
\end{abstract}


\maketitle

\section{Introduction}

Because of the potential for temperature gradients or 
other thermal stresses to cause electronic devices to fail,
knowledge of the coefficients of thermal expansion (CTE) 
of  materials is important.
Either active materials with desired thermal properties or
composites of active and compensating materials can
mitigate the ill effects of
thermal expansion in real devices \cite{Lind12p1125,Chen15p3522}.
A compensating material must be chemically
and electrically compatible with the functional
material and will typically contract with increasing
temperature; i.e.\ it will exhibit negative thermal
expansion (NTE).
While the CTEs for many materials have been cataloged, 
the set of known NTE materials is small, and
materials appropriate for specific applications may 
not yet be available.
A means of rapidly predicting the thermal expansion 
properties of yet-to-be investigated materials will
be extremely useful.

The displacement of the equilibrium positions of atoms in a material
with temperature is the source of thermal expansion.
While anharmonicity in the vibrations of bonded pairs of 
atoms will drive the 
atoms farther apart with increasing temperature, crystal 
structure also plays an essential role in the specific 
thermal expansion characteristics of a material, providing  
behaviors over the range from positive 
thermal expansion (PTE) to NTE.
Accurate theoretical modeling of thermal expansion in materials
necessitates quantum mechanical calculations and
dynamical calculations or, minimally, quasi-harmonic modeling.
As a result, typical calculations of thermal
expansion address the microscopic causes of 
experimentally-observed CTE on a case-by-case basis.
Especially prevalent recently are experimental and
theoretical investigations of
materials that exhibit NTE, such as $\mathrm{ZrW_2O_8}$
\cite{Mary96p90,Pryde96p10973,Ramirez98p4903,Ernst98p1247,%
Mittal99p7234,Mittal01p4692,Cao02p215902,Cao03p014303,%
Hancock04p225501,Tucker05p255501,Gava12p195503,%
Bridges14p045505,Sanson14p3716},
$M_2\mathrm{O}$ (with $M=\mathrm{Cu,Ag,Au}$)
\cite{Gupta14p093507},
$\mathrm{ReO_3}$ \cite{Bozin12p094110,Chatterji09p241902},
and $\mathrm{ScF_3}$.\cite{Li11p195504,Lazar15p224302}
Except for materials with very similar
electronic and structural properties,
a global picture capable of guiding searches for new
materials with desired thermal expansion characteristics is slow
to emerge.
While a useful high-throughput approach to estimating
material thermal properties, employing fits of 
DFT-calculated energy-volume curves to 
equations of state and employing a quasi-harmonic
Debye approximation, has been developed\cite{Toher14p174107}
and recently improved\cite{Toher17p015401},
the approach we present provides an complementary picture
through its focus on local structural properties. 
Comprehensive reviews of thermal expansion in 
materials, emphasizing NTE in both theory and experiment, 
can be found in Refs.~\onlinecite{Barrera05p217} and 
\onlinecite{Dove16p066503}.

For microscopic atomic displacements to create NTE, 
the motions of the atoms must carry them into
spaces already existing within the lattice, while
simultaneously drawing neighboring ions closer together. 
It is well-known\cite{Barrera05p217,Dove16p066503} 
that the definition of the volumetric thermal
expansion $\alpha_v$,
\begin{equation}
 \label{CTE}
 \alpha_v = \frac{1}{V}\left( 
 \frac{\partial V}{\partial T} \right)_P\, ,
\end{equation}
can, with the help of a Maxwell relation, be rewritten as
\begin{equation}
 \label{eq:CTEentropy}
 \alpha_v = -\frac{1}{V} \left(
 \frac{\partial S}{\partial P} \right)_T\, ,
\end{equation}
which shows that for a material to exhibit NTE, entropy
must increase with pressure, contrary to typical 
expectations. 
Typically, decreasing the volume available to a free
particle is associated with decreasing entropy. 
The definition in Eq.~\ref{eq:CTEentropy}
is isothermal, which effectively
means that the momentum space contribution to entropy
changes negligibly relative to the real space contribution.
From this we deduce that applying pressure in an NTE material
effectively increases the volume available to its 
constituent atoms.
On the other hand, applying pressure to PTE materials
(at constant temperature) results in decreasing
entropy, implying a corresponding decrease in the
effective volume available to the atoms that 
follows from the same line of argument used above.
In order for there to be more volume available to
the atoms within a crystalline material 
despite decreasing total volume, atomic 
motions must be directed more significantly 
into the open spaces within the lattice.
In other words, the atomic motions possess
significant components in directions 
perpendicular to the bonds with 
nearest-neighbor atoms.
The difference between NTE and PTE is therefore
directly related to the degree to which ionic thermal 
displacements are longitudinal or transverse 
with respect to the bonds.
Experimental evidence supporting this view is found,
for example,
in a study of NTE in $\mathrm{ScF_3}$, in which
inelastic neutron scattering shows that the
Sc-F bonds lengthen with increasing temperature and that
the material contracts over a wide temperature range
as a result of large transverse motions 
of the F ions \cite{Li11p195504}.
In our recent molecular dynamics investigation,
we demonstrated that thermal expansion in a single
structure, with expanding bonds modeled 
with first- and second-neighbor
interactions via Morse potentials, can be varied from 
NTE to PTE by increasing the second-neighbor interaction strength
relative to the first-neighbor interactions. \cite{Schick16p214304}
By adjusting second-neighbor interactions, we reduced the
transverse motions of the light ions, resulting in
the emergence of PTE in the model.
The key element permitting transverse motion in these
examples is low bond coordination, which is necessarily 
linked to open lattices.

We propose that an approach
to predicting thermal expansion can be found by scanning the
literature for the structures of crystals, focusing
on quantities that may have a relationship to the entropy
of the material and its potential to increase or
decrease with respect to pressure, such as
the space occupied by ions and their bonding coordinations.
We employ the wealth of structural information in databases
such as the  Inorganic Crystal Structure Database 
(ICSD) \cite{Bergerhoff87,Belsky02p364}
and the Crystallographic Open 
Database (COD) \cite{Merkys16p292,Grazulis15p85,%
Grazulis12p420,Grazulis09p726,Downs03p247}
in this work.
By correlating materials with known CTE to their structures,
we can begin to determine
useful descriptors for thermal expansion.
As a result of the lack of complete temperature dependences
of the CTE for many of the materials investigated here,
we focus on structures stable at room temperature and their CTEs.
In Section~\ref{section:proposeddescriptors}, we discuss the 
physical underpinnings for the quantities that we propose
for descriptors of CTE. In Section~\ref{section:results}, we present
the choice of descriptors from the original list of quantities
of interest along with the result of performing a fit using
the descriptors developed.
We conclude with a discussion of the implications this
correlation will have in the search for materials with
desired thermal expansion properties.


\section{\label{section:proposeddescriptors}Proposed descriptors}

For each atom in a unit cell of each material, the number of nearest
neighbors and formal oxidation state are determined. 
From this information we obtain ionic radii \cite{Shannon76p751} 
$r_i$ for each ion and an estimate of the volume
each ion occupies (assuming it occupies
a sphere of volume $V_i = \frac{4}{3} \pi r_i^3$). 
The fractional volume occupied is
the total volume estimated for all the atoms in the unit
cell divided by the volume of the unit cell $V_\mathrm{u.c.}$:
\begin{equation}
\label{eq:fractionalvolume}
v  = \frac{1}{V_\mathrm{u.c.}}{\sum_i V_i}\, .
\end{equation}
A complete list of the data used for this investigation is
provided in the Appendix\ref{section:appendix}.
As seen in Fig.~\ref{fig:alphavolumeplot}, thermal expansion
data for the materials we sampled show that
there is a relationship between thermal expansion
and the volume occupied by atoms in the lattice; more open space
corresponds to a greater likelihood for NTE.

The cluster of points with $\alpha_v \gtrsim 90\times 10^{-6}$/K
 in Fig.~\ref{fig:alphavolumeplot}
consists of binary materials that have the rock salt 
structure, with the corresponding high coordination
providing an explanation for the large positive thermal
expansion in these materials.
We note that Coulomb interactions between second neighbors
will be repulsive and strong.
The data point with the highest volume occupied and
very low PTE belongs to cubic BN.
Because of its zincblende structure, 
the B and N atoms in BN have tetrahedral coordination.
We also note that BN has a high bulk modulus, which
implies that the tetrahedral bonds are
strong and maintain their directionality. 
Although the volume occupancy of BN is high,
the strong bonds produce a smaller thermal
expansion.
(The situation in BN is reminiscent of the diamond
phases of C, Si, and Ge, which all have small values
of $\alpha_v$
at room temperature. It is notable that Si exhibits
NTE at lower temperatures, indicating a lower barrier
to transverse motions of its atoms.)
Although low bond coordination
in a material is a possible descriptor to predict 
thermal expansion coefficients, it is not sufficient
on its own.  Low coordination, such as seen in BN, 
is a result of directional bonding. But 
low coordination also
occurs at the linking atoms in
a crystal structure that is composed of
groups of atoms forming stable units, as is the
case for ScF$_3$, ReO$_3$ and other perovskite materials.
Collective rotational-vibrational motion of these units
is known as a rigid unit mode (RUM) \cite{Hammonds96p1057}.
RUMs are cited as the underlying cause of NTE in ScF$_3$,
ReO$_3$ \cite{Chatterji09p241902}, and, arguably, 
in ZrW$_2$O$_8$ \cite{Tucker05p255501}.
The atoms at the linking points of the rigid units
typically are two-fold coordinated, while the atoms
at the centers are more highly coordinated.
The thermal vibrations of the RUM tend to be low
frequency, carrying low-coordination 
atoms significant distances in directions 
transverse to the bonds.
We note that the RUM is also presented as a foundational
model for displacive phase transitions \cite{Dove97p213}
connecting these vibrational modes to symmetry changes
at phase boundaries.
The importance of phase boundaries for which the 
higher-temperature phase has a smaller 
volume (and higher entropy) 
has been introduced as a thermodynamic cause of 
NTE \cite{Liu11p664,Liu14p7043}.  As the temperature
is increased toward the phase boundary from below,
thermal fluctuations introducing an increasing fraction
of the lower-volume, higher-entropy structure into the 
lattice is shown to be the necessary condition for NTE
\cite{Liu11p664,Liu14p7043}.
This is consistent with the observation above that
greater entropy must be associated with smaller volume
in NTE materials.
As a result of these observations, 
we suggest that the variability of bonding coordinations
in a material, indicative of rigid units, is a
key structural quantity to be used as a descriptor
for thermal expansion.

\begin{figure}
\includegraphics[width=3.4in]{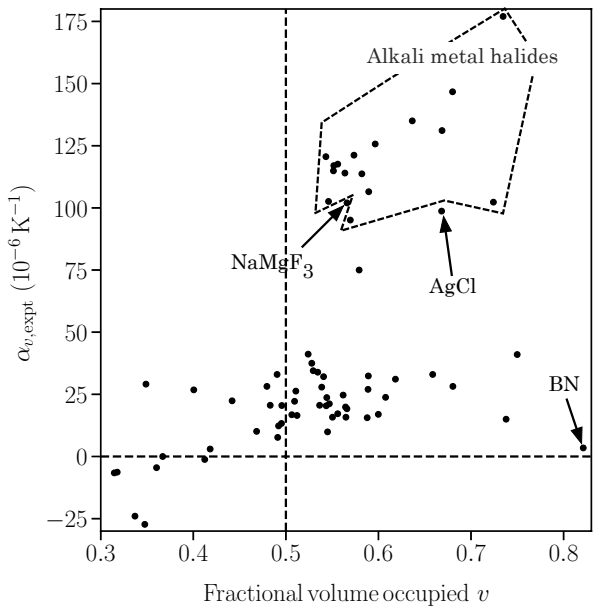}%
\caption{\label{fig:alphavolumeplot} The 
experimental coefficient of thermal expansion 
($\alpha_{v,\mathrm{expt}}$) 
is plotted as a function of fractional volume occupied as
defined in Eq.~\ref{eq:fractionalvolume} for the
materials gleaned from the literature.  Apparently,
NTE is possible only when $v$ is less than $\approx 0.5$.}
\end{figure}

\section{\label{section:results}Results and discussion}

\begin{figure}
\includegraphics[width=3.5in]{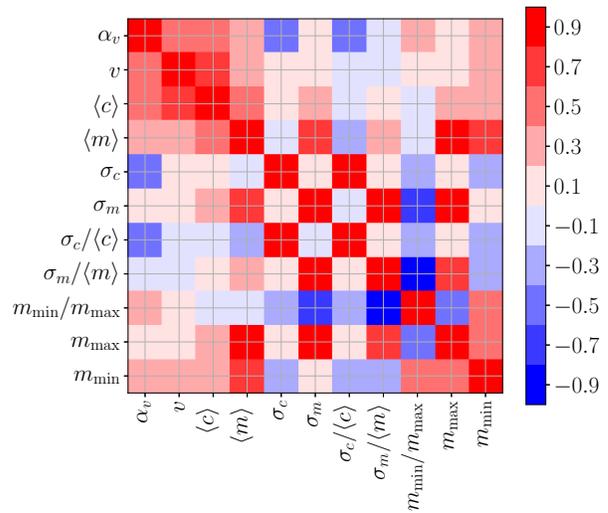}%
\caption{\label{fig:correlation}
(Color online)
The correlations (Pearson-$r$) between candidate 
descriptors and the coefficient of thermal expansion
$\alpha_v$
for 71 materials taken from the literature.
Moderate linear correlation with 
$\alpha_v$ is found
for both the mean coordination and occupied volume. 
Weak-to-moderate correlation is exhibited by
both the mean mass and the ratio of minimum to 
maximum atomic mass. Focusing on quantities
which correlate strongly with $\alpha_v$ leads to
a set of three descriptors: the occupied volume,
the mean coordination, and standard deviation of the 
coordination.}
\end{figure}

Quantities extracted in the search for descriptors,
in addition to $v$,
are the mean bond coordination $\langle c \rangle$,
its standard deviation $\sigma_c$, the mean atomic mass 
$\langle m \rangle$, the minimum mass 
$m_\mathrm{min}$, the maximum mass $m_\mathrm{max}$,
and the standard deviation of mass $\sigma_m$.
In Fig.~\ref{fig:correlation}, we display the linear 
correlations between descriptor candidates
and the experimental $\alpha_v$. In addition, we
include unitless ratios $\sigma_c/\langle c \rangle$,
$\sigma_m/\langle m\rangle$, and the ratio of
minimum to maximum mass $m_\mathrm{min}/m_\mathrm{max}$.
The quantities that have the strongest
correlations with the $\alpha_v$ are $v$ 
and $\langle c \rangle$, 
with correlations 0.48 and 0.55, respectively. 
The next-most-highly correlated quantity is
$\sigma_c$, with a (negative) 
correlation of $-0.42$ to $\alpha_v$. Furthermore,
this quantity does not correlate with either
the volume ratio $v$ or the mean coordination 
$\langle c \rangle$.
The candidates involving mass have low correlations
with $\alpha_v$ and are neglected.
Finally, we determine that 
$v$, $\sigma_c$, and $\langle c \rangle$
are a good set of descriptors.

\begin{figure}
 \includegraphics[width=3.4in]{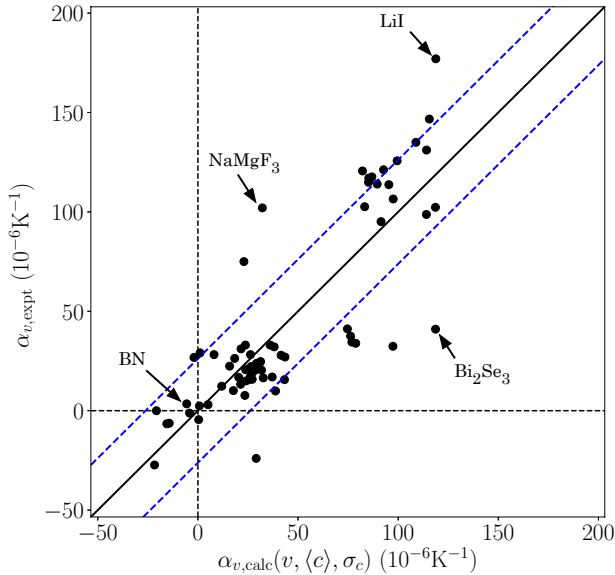}%
\caption{\label{fig:multi-fit} (Color online)
Correlation plot of room-temperature 
experimental and calculated $\alpha_v$
generated by a ten-parameter second-order polynomial fit
to experimental data using Eq.~\ref{eq:fit} with the
parameters in Table~\ref{table:coefficients}.  
The root-mean-squared deviation of the calculated 
coefficients of expansion, indicated with blue
dashed lines, is $26\times 10^{-6} \mathrm{K}^{-1}$.
The Pearson-$r$ for this fit is $0.83$.}
\end{figure}

To discover the relationships between these descriptors
and the experimentally-measured values of $\alpha_v$,
we performed a second-order polynomial fit of these three
descriptors to the 71 materials in our data sample, using the
relation
\begin{equation}
\label{eq:fit}
\alpha_{v,\mathrm{calc}} \left( 
v , \langle c \rangle , 
\sigma_c
\right) = 
\sum_{i+j+k\le 2}
c_{ijk}\, v^i \langle c \rangle^j
\left(\sigma_c\right)^k\, .
\end{equation}
The correlation plot for this fit is displayed in
Fig.~\ref{fig:multi-fit}, and the coefficients are listed
in Table~\ref{table:coefficients}.
The correlation coefficient is 0.83, and the RMS-deviation
between the fit and the data is $26\times 10^{-6}~\mathrm{K}^{-1}$,
a noticeable improvement in correlation over the data displayed in 
Fig.~\ref{fig:alphavolumeplot}, with outlying (alkali-halide) 
points more closely fit.  
Including only linear terms in the fit did not produce an
acceptable interpolation.  Adding cubic terms produced
a substantially better fit but the trends we describe below
are unchanged by the additional terms.
Points to the right of the diagonal are ones for 
which the model predicts thermal expansion greater
than the experimentally measured value.
The previously noted deviation in BN
has been nearly eliminated.  The predicted
values for the ionic solids have also been increased toward
their experimental values.
The material with largest positive deviation 
($\alpha_{v,\mathrm{calc}} > \alpha_{v,\mathrm{expt}}$) 
is $\mathrm{Bi_2Se_3}$.   
The largest negative deviation arises for $\mathrm{LiI}$.
\begin{table}
\caption{\label{table:coefficients}The parameters
used in Eq.~\ref{eq:fit} to evaluate 
$\alpha_{v,\mathrm{calc}} (v,\langle c \rangle, \sigma_c)$. Uncertainties
are based on a nominal experimental $\alpha_v$ 
error of $10^{-6}$~K$^{-1}$.}
\begin{ruledtabular}
\begin{tabular}{cccc}
$i$&$j$&$k$&$c_{ijk}$\\
\hline
0&0&0&$-290 \pm 5$\\
1&0&0&$685 \pm 9$\\
0&1&0&$15 \pm 1$\\
0&0&1&$103.7 \pm 0.8$\\
1&1&0&$123 \pm 1$\\
1&0&1&$-132 \pm 1$\\
0&1&1&$-14.3 \pm 0.1$\\
2&0&0&$-964 \pm 7$\\
0&2&0&$-5.7 \pm 0.2$\\
0&0&2&$12.0 \pm 0.1$\\
\end{tabular}
\end{ruledtabular}
\end{table}

\begin{figure}
 \includegraphics[width=3.7in]{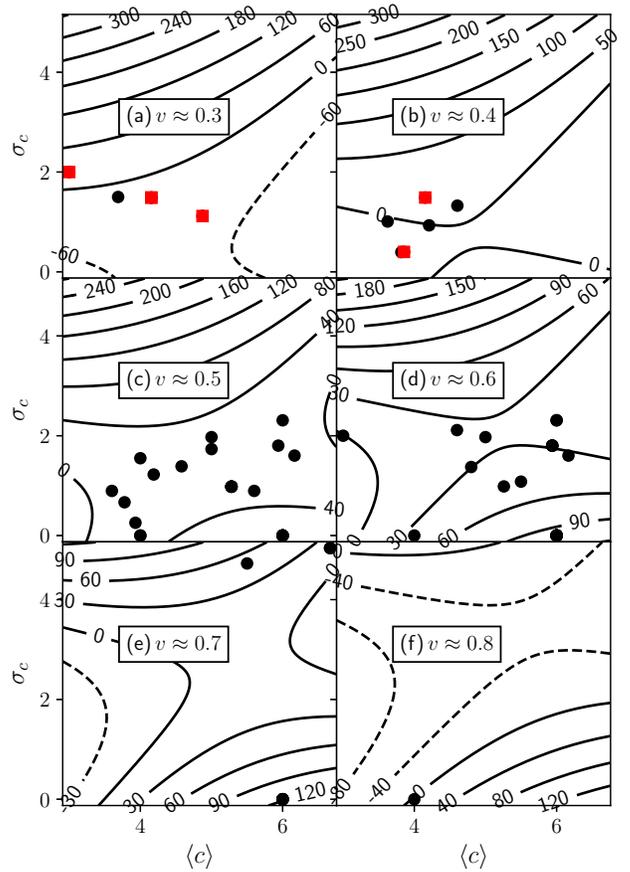}%
\caption{\label{fig:layers} (Color online)
Contour plots showing the dependence of the calculated 
$\alpha_{v}$ upon $\langle c\rangle$ 
and $\sigma_c$ at fixed volume ratios $v$, as indicated. 
The markers plotted display the descriptor values
for the materials used in the fit, with black circles for
PTE materials and red squares for NTE materials. 
The volume ratios for the data points displayed 
in each panel fall within $\pm 0.05$ of the
value of $v$ indicated in each panel.
All experimental $\alpha_v$ values are measured
at room temperature.}
\end{figure}

In order to visualize the model and to isolate ranges
of the descriptors that characterize
the CTE, we present contours of constant 
$\alpha_{v,\mathrm{calc}}$ across the ranges of
$\langle c\rangle$ and $\sigma_c$
at fixed volume ratios $v=0.3, 0.4, \dots, 0.8$ in 
Fig.~\ref{fig:layers}. Superimposed on the contours
are points indicating the descriptor values of data
used in the fit, including volume ratios within $\pm 0.05$
of the value of $v$ displayed in the corresponding panel.
NTE materials
are indicated in Fig.~\ref{fig:layers} with red dots.
As noted previously, materials with the lowest filled volume
ratios are most likely to have NTE.  
In panels (a) and (b) in Fig.~\ref{fig:layers} we see 
that the NTE materials generally have 
$\langle c \rangle \lesssim 4$
and coordination deviations  $\sigma_c \approx 2$,
the exception being LiAlSiO$_4$, which has 
$\alpha_{v,\mathrm{expt}}=1.2\times 10^{-6}~\mathrm{K}^{-1}$.
In addition to open space, important contributors to
NTE are low mean coordination and a distribution
of local coordinations.

The majority of materials in this work have
volume ratios around 0.5 to 0.6 and 
are displayed in panels (c) and (d) in 
Fig.~\ref{fig:layers}. 
These materials also generally possess 
$\alpha_{v,\mathrm{expt}} \approx 30\times 10^{-6}~\mathrm{K}^{-1}$ 
that would be considered typical.
Mean coordinations are generally greater than that
found in the NTE materials in panels (a) and (b).
Many of the materials in this volume ratio neighborhood
have zero coordination variance as a result of their
simple structures. In particular,
the ions in the rock-salt-structured alkali halides 
have identical coordination environments and 
are highly coordinated, resulting in the highest values of
PTE.
Most of the perovskite oxides in this study
also appear in panels (c) and (d) of Fig.~\ref{fig:layers}
and have greater mean coordinations but they
have greater deviations in coordination, due to
a structure that consists of octahedra linked by
two-fold coordinated oxygen atoms. This structure
leads to a greater ability for atomic motions
that are transverse to the bond and, therefore, to
lower values of PTE. 
In panels (e) and (f) of Fig.~\ref{fig:layers} the 
data points occupy the highest and lowest values
of $\sigma_c$.  A number of the samples have identical
values of both $\langle c \rangle = 6$ and $\sigma_c = 0$,
similar to panels (c) and (d), again leading
to larger PTE.

Summarizing, the trends indicate that NTE requires
$v \lesssim 0.5$, 
$\langle c \rangle \lesssim 4$, and 
$\sigma_c \approx 2$, which is a large
value in comparison to the mean coordination.
With larger volume ratios, again it is observed that
materials with $\sigma_c$ around half the value 
of $\langle c \rangle$ have the lowest thermal 
expansion values. These materials can typically 
also be modeled within the RUM \cite{Hammonds96p1057}.
Highly coordinated materials, 
$\langle v \rangle \approx 6$,
with zero deviation generally have strong PTE.

The CTE of $\mathrm{Bi_2Se_3}$ is overestimated,
by the present picture. While the
descriptor values, $v=0.75$, $\langle c \rangle=6$,
and $\sigma_c=0$, would predict large PTE,
the material has a layered structure, leading to
more open space and to anisotropic expansion, 
unlike most of the materials in this work.
We note that BN, which was a significant outlier in 
Fig.~\ref{fig:alphavolumeplot},
is now slightly underestimated, with 
$\alpha_{v,\mathrm{calc}}
= -5.6\times 10^{-6}~\mathrm{K}^{-1}$.
A number of materials with zincblende or diamond
lattice structures, such as Si, show NTE for a range of low 
temperatures \cite{Sparks67p779}.  A relatively low resistance
for B-N-B (or Si-Si-Si) bond angles to change may account for
the discrepancy between the model and experiment. While the 
present model does not have a parameter that directly
represents this particular characteristic, the mean 
coordination does provide some insight into it.
The $\alpha_v$ values of LiI and NaMgF$_3$ 
are underestimated in the the present picture.
These outliers include atoms from the top row of 
the periodic table, suggesting
this as a reason for the deviations.

\section{\label{section:conclusion}Conclusion}
The thermal expansion properties of solids emerge from the 
dynamics of ions and the bonds that join them.
While the full quantum-mechanical nature of bonding  
and atomic arrangement together are responsible for this, 
we demonstrate that significant insight into the 
value of the coefficient of thermal expansion at room 
temperature is gained by focusing on the structural 
aspects of a material.
Descriptors based on the open space, as quantified
by ionic volume and unit cell volume, mean bond
coordination, and variability of local
bond coordination permit estimation of
the thermal expansion coefficient of a variety of materials,
including ionic binary compounds, perovskites, silicates, and
other oxides, using only the crystal structure of 
the material and the ionic radii.
A higher value of the coefficient of thermal expansion
is expected when a material has a high ionic volume
filling ratio and single-valued local
coordination, especially when the value is large.
On the other hand, negative values of the thermal
expansion coefficient arise in materials for which the
volume filling is low and the mean local bonding coordination
is small, coupled with a deviation of coordination that
is large in comparison to its mean value.
These descriptors not only reinforce the physical description
thermal expansion but also provide guidance for 
the search for new materials capable of mitigating 
thermal expansion in devices.

\appendix*
\section{\label{section:appendix}}
Table~\ref{table:alldata} lists the sources of structures and
CTE used in this work. The measurements are taken from
available room-temperature data.
\begin{widetext}
\LTcapwidth=\columnwidth
\begin{longtable}{lccrrccc}
\caption{\label{table:alldata}
All structural data used in this work are found in either
the Crystal Online Database (COD) or in the Inorganic Crystal Structure
Database (ICSD). Each material is listed with its database index. 
Experimental $\alpha_v$ values are listed with their data sources 
indicated. The volume ratio $v$, mean coordination 
$\langle c \rangle$, and standard deviation of
the coordination $\sigma_c$ 
are evaluated as described in the text.} \\
\hline \hline \\[0.5ex]
   \multicolumn{1}{c}{{Material}} &
   \multicolumn{1}{c}{{COD}} &
   \multicolumn{1}{c}{{ICSD}} &
   \multicolumn{1}{c}{{$\alpha_{v,\mathrm{expt}}$}} &
   \multicolumn{1}{c}{{$\alpha_{v,\mathrm{calc}}$}} &
   \multicolumn{1}{c}{$v$} &
   \multicolumn{1}{c}{$\langle c \rangle$} &
   \multicolumn{1}{c}{$\sigma_c$} \\ [0.2ex] 
   \hline
   \\[0.5ex]
\endfirsthead

\multicolumn{3}{c}{{\tablename} \thetable{} -- Continued} \\[0.5ex]
   \hline \hline \\[0.5ex]
   \multicolumn{1}{c}{{Material}} &
   \multicolumn{1}{c}{{COD}} &
   \multicolumn{1}{c}{{ICSD}} &
   \multicolumn{1}{c}{{$\alpha_{v,\mathrm{expt}}$}} &
   \multicolumn{1}{c}{{$\alpha_{v,\mathrm{calc}}$}} &
   \multicolumn{1}{c}{$v$} &
   \multicolumn{1}{c}{$\langle c \rangle$} &
   \multicolumn{1}{c}{$\sigma_c$} \\ [0.2ex]
   \hline
   \\[0.5ex]
\endhead

  \multicolumn{3}{l}{{Continued\ldots}} \\ [0.1ex]
\endfoot

  \\
  \hline \hline
\endlastfoot
$\mathrm{AgBr}$   &  $1509151$  &    &  $102.30$\cite{Kumar59p364}  &  $118.61$  &  $0.72$   &  $6.00$  &  $0.00$  \\ [0.1ex] 
$\mathrm{AgCl}$   &    &  $56538$  &  $98.70$\cite{Kumar59p364}  &  $114.10$  &  $0.67$   &  $6.00$   &  $0.00$  \\ [0.1ex] 
$\mathrm{Al_2FeO_4}$   &  $9001970$  &    &  $15.60$\cite{Fei95p29}  &  $43.22$  &  $0.59$   &  $5.26$  &  $0.98$  \\ [0.1ex] 
$\mathrm{AlN}$   &    &  $54697$  &  $7.68$\cite{Slack75p89}  &  $23.45$  &  $0.49$   &  $4.00$   &  $0.00$  \\ [0.1ex] 
$\mathrm{Al_2Fe_3Si_3O_{12}}$   &  $1531283$  &    &  $15.80$\cite{Fei95p29}  &  $27.12$  &  $0.56$   &  $5.94$  &  $1.80$  \\ [0.1ex] 
$\mathrm{Al_2Mg_3Si_3O_{12}}$   &  $9000138$  &    &  $19.90$\cite{Fei95p29}  &  $27.08$  &  $0.56$   &  $5.94$  &  $1.80$  \\ [0.1ex] 
$\mathrm{Al_2Mn_3Si_3O_{12}}$   &  $9002695$  &    &  $17.20$\cite{Fei95p29}  &  $26.26$  &  $0.56$   &  $5.94$  &  $1.80$  \\ [0.1ex] 
$\mathrm{Al_2Si_5O}$   &  $1011204$  &    &  $13.29$\cite{Fei95p29}  &  $21.33$  &  $0.50$   &  $4.19$  &  $1.22$  \\ [0.1ex] 
$\mathrm{Al_2W_3O_{12}}$   &    &  $90936$  &  $-4.50$\cite{Woodcock99p2508}  &  $0.38$  &  $0.36$   &  $4.15$   &  $1.49$  \\ [0.1ex] 
$\mathrm{BN}$   &  $9008834$  &    &  $3.45$\cite{Slack75p89}  &  $-5.60$  &  $0.82$   &  $4.00$  &  $0.00$  \\ [0.1ex] 
$\mathrm{BPO_4}$   &  $1010299$  &    &  $28.20$\cite{Achary04p3918}  &  $8.06$  &  $0.48$   &  $3.60$  &  $0.89$  \\ [0.1ex] 
$\mathrm{BeAl_2O_4}$   &  $9000120$  &    &  $23.80$\cite{Fei95p29}  &  $29.25$  &  $0.61$   &  $4.80$  &  $1.37$  \\ [0.1ex] 
$\mathrm{BeO}$   &    &  $15620$  &  $16.95$\cite{Slack75p89}  &  $37.08$  &  $0.60$   &  $4.00$   &  $0.00$  \\ [0.1ex] 
$\mathrm{Be_2SiO_4}$   &  $9001088$  &    &  $16.80$\cite{Fei95p29}  &  $20.35$  &  $0.51$   &  $3.93$  &  $0.26$  \\ [0.1ex] 
$\mathrm{Be_3Al_2Si_6O_{18}}$   &  $1010541$  &    &  $3.00$\cite{Megaw71p1007}  &  $5.06$  &  $0.42$   &  $4.21$  &  $0.93$  \\ [0.1ex] 
$\mathrm{Bi_2Se_3}$   &  $1530736$  &    &  $41.01$\cite{Chen11p261912}  &  $118.67$  &  $0.75$   &  $6.00$  &  $0.00$  \\ [0.1ex] 
$\mathrm{CaGeO_3}$   &  $9000904$  &    &  $31.10$\cite{Fei95p29}  &  $21.53$  &  $0.62$   &  $6.00$  &  $2.31$  \\ [0.1ex] 
$\mathrm{CaOTi_3}$   &  $1000022$  &    &  $75.00$\cite{Megaw71p1007}  &  $22.95$  &  $0.58$   &  $6.00$  &  $2.31$  \\ [0.1ex] 
$\mathrm{CaO}$   &  $1000044$  &    &  $37.50$\cite{Kumar59p364}  &  $76.25$  &  $0.53$   &  $6.00$  &  $0.00$  \\ [0.1ex] 
$\mathrm{Ca_3Al_2Si_3O_{12}}$   &  $9000236$  &    &  $19.20$\cite{Isaak92p106}  &  $27.21$  &  $0.57$   &  $5.94$  &  $1.80$  \\ [0.1ex] 
$\mathrm{Ca_3Fe_2Si_3O_{12}}$   &  $9007693$  &    &  $20.60$\cite{Fei95p29}  &  $23.82$  &  $0.54$   &  $5.94$  &  $1.80$  \\ [0.1ex] 
$\mathrm{Cr_2FeO_4}$   &  $9007325$  &    &  $9.90$\cite{Fei95p29}  &  $38.76$  &  $0.55$   &  $5.28$  &  $0.98$  \\ [0.1ex] 
$\mathrm{Cr_2MgO_4}$   &  $9006180$  &    &  $16.50$\cite{Fei95p29}  &  $32.62$  &  $0.51$   &  $5.28$  &  $0.98$  \\ [0.1ex] 
$\mathrm{FeO}$   &  $1011169$  &    &  $33.90$\cite{Fei95p29}  &  $78.82$  &  $0.53$   &  $6.00$  &  $0.00$  \\ [0.1ex] 
$\mathrm{FeTiO_3}$   &  $1011033$  &    &  $27.90$\cite{Fei95p29}  &  $41.52$  &  $0.54$   &  $5.60$  &  $0.89$  \\ [0.1ex] 
$\mathrm{Fe_3O_4}$   &  $9002316$  &    &  $20.60$\cite{Fei95p29}  &  $25.50$  &  $0.48$   &  $5.28$  &  $0.98$  \\ [0.1ex] 
$\mathrm{GaN}$   &  $1010168$  &    &  $10.11$\cite{Iwanaga00p2451}  &  $17.70$  &  $0.47$   &  $4.00$  &  $0.00$  \\ [0.1ex] 
$\mathrm{GeZn_2O_4}$   &    &  $16173$  &  $0.00$\cite{Stevens04p349}  &  $-20.73$  &  $0.37$   &  $3.82$   &  $0.39$  \\ [0.1ex] 
$\mathrm{HfO_2}$   &  $1528988$  &    &  $15.80$\cite{Fei95p29}  &  $26.12$  &  $0.55$   &  $6.17$  &  $1.60$  \\ [0.1ex] 
$\mathrm{KBr}$   &  $9008650$  &    &  $117.60$\cite{Pathak73p477}  &  $86.85$  &  $0.56$   &  $6.00$  &  $0.00$  \\ [0.1ex] 
$\mathrm{KCl}$   &    &  $165593$  &  $114.90$\cite{Kumar59p364}  &  $85.20$  &  $0.55$   &  $6.00$   &  $0.00$  \\ [0.1ex] 
$\mathrm{KF}$   &  $9008652$  &    &  $95.10$\cite{Rapp73p3919}  &  $91.44$  &  $0.57$   &  $6.00$  &  $0.00$  \\ [0.1ex] 
$\mathrm{KI}$   &  $9008654$  &    &  $113.70$\cite{Pathak75p155}  &  $95.34$  &  $0.58$   &  $6.00$  &  $0.00$  \\ [0.1ex] 
$\mathrm{KNbO_3}$   &  $1531431$  &    &  $15.00$\cite{Megaw71p1007}  &  $24.05$  &  $0.74$   &  $5.50$  &  $4.73$  \\ [0.1ex] 
$\mathrm{LiAlSiO_4}$   &  $9002541$  &    &  $-1.20$\cite{Fei95p29}  &  $-4.16$  &  $0.41$   &  $3.85$  &  $0.40$  \\ [0.1ex] 
$\mathrm{LiAlSi_2O_6}$   &  $9000347$  &    &  $22.20$\cite{Fei95p29}  &  $26.62$  &  $0.51$   &  $4.58$  &  $1.39$  \\ [0.1ex] 
$\mathrm{LiBr}$   &  $9008664$  &    &  $146.70$\cite{Rapp73p3919}  &  $115.56$  &  $0.68$   &  $6.00$  &  $0.00$  \\ [0.1ex] 
$\mathrm{LiCl}$   &  $9008665$  &    &  $131.10$\cite{Rapp73p3919}  &  $114.16$  &  $0.67$   &  $6.00$  &  $0.00$  \\ [0.1ex] 
$\mathrm{LiF}$   &    &  $181799$  &  $106.50$\cite{Pathak72p30}  &  $97.51$  &  $0.59$   &  $6.00$   &  $0.00$  \\ [0.1ex] 
$\mathrm{LiI}$   &  $9008669$  &    &  $177.00$\cite{Kumar59p364}  &  $118.79$  &  $0.73$   &  $6.00$  &  $0.00$  \\ [0.1ex] 
$\mathrm{MgF_2}$   &  $1526229$  &    &  $33.00$\cite{Megaw71p1007}  &  $23.71$  &  $0.49$   &  $4.00$  &  $1.55$  \\ [0.1ex] 
$\mathrm{MgFe_2O_4}$   &  $1011245$  &    &  $20.50$\cite{Fei95p29}  &  $28.78$  &  $0.50$   &  $5.28$  &  $0.98$  \\ [0.1ex] 
$\mathrm{MgGeO_3}$   &  $9000957$  &    &  $22.40$\cite{Fei95p29}  &  $15.82$  &  $0.44$   &  $4.61$  &  $1.33$  \\ [0.1ex] 
$\mathrm{MgO}$   &    &  $9863$  &  $32.40$\cite{Suzuki75p145}  &  $97.40$  &  $0.59$   &  $6.00$   &  $0.00$  \\ [0.1ex] 
$\mathrm{Mg_2GeO_4}$   &  $9010486$  &    &  $32.10$\cite{Fei95p29}  &  $38.08$  &  $0.54$   &  $5.28$  &  $0.98$  \\ [0.1ex] 
$\mathrm{MnO}$   &  $1010393$  &    &  $34.50$\cite{Fei95p29}  &  $76.88$  &  $0.53$   &  $6.00$  &  $0.00$  \\ [0.1ex] 
$\mathrm{Mo_3Fe_2O_{12}}$   &  $1524203$  &    &  $29.10$\cite{Tyagi02p207}  &  $1.06$  &  $0.35$   &  $3.69$  &  $1.50$  \\ [0.1ex] 
$\mathrm{NaAlSi_2O_6}$   &  $9000143$  &    &  $24.70$\cite{Fei95p29}  &  $31.29$  &  $0.56$   &  $5.00$  &  $1.97$  \\ [0.1ex] 
$\mathrm{NaAlSi_3O_8}$   &  $9000526$  &    &  $26.80$\cite{Fei95p29}  &  $-1.99$  &  $0.40$   &  $3.62$  &  $1.01$  \\ [0.1ex] 
$\mathrm{NaBr}$   &  $9007464$  &    &  $125.70$\cite{Rapp73p3919}  &  $99.48$  &  $0.60$   &  $6.00$  &  $0.00$  \\ [0.1ex] 
$\mathrm{NaCl}$   &  $4320809$  &    &  $121.20$\cite{Pathak70p655}  &  $92.71$  &  $0.57$   &  $6.00$  &  $0.00$  \\ [0.1ex] 
$\mathrm{NaCrSi_2O_6}$   &  $1531195$  &    &  $20.40$\cite{Fei95p29}  &  $31.77$  &  $0.54$   &  $5.00$  &  $1.97$  \\ [0.1ex] 
$\mathrm{NaF}$   &    &  $262837$  &  $102.60$\cite{Pathak73p477}  &  $83.26$  &  $0.55$   &  $6.00$   &  $0.00$  \\ [0.1ex] 
$\mathrm{NaI}$   &  $9008681$  &    &  $135.00$\cite{Rapp73p3919}  &  $108.86$  &  $0.64$   &  $6.00$  &  $0.00$  \\ [0.1ex] 
$\mathrm{NaMgF_3}$   &  $9001612$  &    &  $102.00$\cite{Megaw71p1007}  &  $32.18$  &  $0.57$   &  $4.60$  &  $2.11$  \\ [0.1ex] 
$\mathrm{NaNbO_3}$   &  $1011064$  &    &  $33.00$\cite{Megaw71p1007}  &  $36.08$  &  $0.66$   &  $6.67$  &  $5.03$  \\ [0.1ex] 
$\mathrm{ReO_3}$   &   &  $77679$  &  $2.4$\cite{Chatterji09p241902}  &  $0.73$  &  $0.60$   &  $3.00$  &  $2.00$  \\ [0.1ex] 
$\mathrm{RbBr}$   &  $9008706$  &    &  $117.00$\cite{Pathak73p477}  &  $85.33$  &  $0.55$   &  $6.00$  &  $0.00$  \\ [0.1ex] 
$\mathrm{RbCl}$   &    &  $22166$  &  $120.60$\cite{Srivastava73p2069}  &  $82.19$  &  $0.54$   &  $6.00$   &  $0.00$  \\ [0.1ex] 
$\mathrm{RbI}$   &  $9008710$  &    &  $114.00$\cite{Pathak75p155}  &  $89.51$  &  $0.56$   &  $6.00$  &  $0.00$  \\ [0.1ex] 
$\mathrm{ScAlO_3}$   &  $9005883$  &    &  $27.00$\cite{Fei95p29}  &  $43.45$  &  $0.59$   &  $5.50$  &  $1.08$  \\ [0.1ex] 
$\mathrm{ScF_3}$   &    &  $261072$  &  $-24.00$\cite{Greve10p15496}  &  $29.10$  &  $0.34$   &  $3.00$   &  $2.00$  \\ [0.1ex] 
$\mathrm{Sc_2Mo_3O_{12}}$   &    &  $391467$  &  $-6.30$\cite{Evans97p580}  &  $-14.32$  &  $0.32$   &  $4.15$   &  $1.49$  \\ [0.1ex] 
$\mathrm{Sc_2W_3O_{12}}$   &    &  $50941$  &  $-6.60$\cite{Evans97p580}  &  $-15.54$  &  $0.31$   &  $4.15$   &  $1.49$  \\ [0.1ex] 
$\mathrm{SrO}$   &    &  $163625$  &  $41.16$\cite{Kumar59p364}  &  $74.65$  &  $0.52$   &  $6.00$   &  $0.00$  \\ [0.1ex] 
$\mathrm{SrTiO_3}$   &  $7212245$  &    &  $28.20$\cite{deLigny96p3013}  &  $26.26$  &  $0.68$   &  $6.67$  &  $5.03$  \\ [0.1ex] 
$\mathrm{SrZrO_3}$   &  $1521387$  &    &  $26.30$\cite{deLigny96p3013}  &  $18.34$  &  $0.51$   &  $6.00$  &  $2.31$  \\ [0.1ex] 
$\mathrm{TiO_2}$   &  $4102355$  &    &  $23.70$\cite{Megaw71p1007}  &  $30.64$  &  $0.54$   &  $5.00$  &  $1.73$  \\ [0.1ex] 
$\mathrm{ZrO_2}$   &  $2300544$  &    &  $21.20$\cite{Fei95p29}  &  $25.53$  &  $0.55$   &  $6.17$  &  $1.60$  \\ [0.1ex] 
$\mathrm{ZrSiO_4}$   &  $1011265$  &    &  $12.30$\cite{Fei95p29}  &  $11.87$  &  $0.49$   &  $3.78$  &  $0.67$ \\ [0.1ex]
$\mathrm{ZrW_2O_8}$ &  & $262061$ & $-27.30$\cite{Evans96p2809} & $-21.68$ & $0.35$  & $4.87$  & $1.12$  \\ [0.1ex]
\end{longtable}
\end{widetext}

%

\begin{acknowledgments}
This work has been supported by the Department of Energy Office of 
Basic Energy Sciences, under grant number DE-FG02-07ER46431.
A.M.R. acknowledges support from the Office of Naval Research 
under grant number N00014-17-1-2574.
Computational support was provided by the National Energy
Research Scientific Computing Center (NERSC).
\end{acknowledgments}

%

\end{document}